\documentclass[a4paper,11pt]{article}
\pdfoutput=1 

\usepackage{jinstpub} 

\usepackage{lineno}

\usepackage{booktabs, multirow} 
\usepackage{soul}
\usepackage[table]{xcolor} 
\usepackage{changepage,threeparttable} 

\title{\boldmath 
Performance Evaluation of an RFSoC Operating in a 1.25 Tesla Magnetic Field
}

\author{L. Ruckman,}
\author{A. Dragone,}
\author{and R. Herbst}
\affiliation{
    SLAC National Accelerator Laboratory,
    2575 Sand Hill Road, M/S 96,
    Menlo Park, CA 94025, USA
}

\emailAdd{ruckman@slac.stanford.edu}

\abstract{

\noindent

The Belle II experiment at the SuperKEKB collider is preparing for an upgrade after 2027-2028 to handle higher luminosity and increased background rates. A Radio Frequency System-on-Chip (RFSoC) has been identified as a potential candidate for a common front-end upgrade for subsystems requiring high-speed waveform digitization. The RFSoC's ADC and DAC channels were tested across various magnetic field strengths and a few different field orientations. Power consumption and boot memory functionality were also assessed. Results indicate stable operation with negligible performance degradation, suggesting the RFSoC's viability for high-speed digitization tasks in high magnetic field environments.

}


\begin{document}
\maketitle
\flushbottom
\setlength{\parskip}{0pt}


\section{Background and Motivation}
\label{sec:Background}

The Belle II experiment at the SuperKEKB \( e^{+}e^{-} \) collider in Japan is designed to investigate potential deviations from the Standard Model through precise measurements of B meson decays \cite{belle2KEK}. These measurements provide valuable insights into new physics beyond the Standard Model. A tentative schedule for Belle II's Long Shutdown \#2 (LS2) has been set for the period after 2027–2028 to allow for multiple subsystem upgrades to the Belle II detector \cite{belle2KEK_upgrade}. By upgrading these subsystem, Belle II can maintain the precision required to detect rare events and phenomena potentially outside the Standard Model, helping to unlock new insights in particle physics. 

These upgrades are required due to the inability of some Belle II subsystems to meet the demands of higher luminosity and increased background rates expected post-LS2 \cite{belle2KEK_snomass2022}. To achieve greater precision and data handling capacity, the upgrades aim to enhance the performance and resilience of key Belle II subsystems. Without these improvements, critical components would struggle to maintain sensitivity and accuracy under the elevated collision rates and background noise anticipated with the planned luminosity increase.

Each subsystem of the Belle II detector uses different designs of Application-Specific Integrated Circuits (ASICs), commercial Analog-to-Digital Converters (ADCs), or commercial Time-To-Digital Converters (TDCs) for their respective signal digitization. The lack of commonality in the designs of different sub-detector front ends, apart from the timing system, causes the Non-Recurring Engineering (NRE) costs for development and operational maintenance to be higher than those for a more modular design. The sub-detector upgrades during LS2 provide an opportunity to consider adopting a more modular approach for two or more subsystems.

Two subsystems have been identified as potential candidates for a common front-end upgrade due to their fundamental function as high-speed gigasample per second (GSPS) waveform digitizers: iTOP (Imaging Time-of-Propagation) and KLM (K-Long Muon). iTOP reads 8,192 channels using multiple 8-channel waveform sampling ASICs called the 'Ice Ray Sampler Version X' (IRSX), sampling at 2.714 GSPS \cite{iTOP}. KLM reads approximately 20,000 scintillator channels using multiple 16-channel waveform sampling ASICs called the 'TeV Array Readout Electronics with GSPS Sampling and Event Trigger Version X' (TARGETX), sampling at 1 GSPS \cite{KLM}.

The Radio Frequency System-on-Chip (RFSoC) developed by AMD has been recognized as a significant advancement in the field of radio frequency and digital signal processing. The RFSoC integrates data converters and programmable logic onto a single package which is designed to streamline the development of high-performance radio frequency systems \cite{WP489,rfsoc}. The third generation ("Gen 3") of RFSoCs can directly RF sample up to 5 GSPS with 8 channels or up to 2.5 GSPS with 16 channels, making it an ideal commercial solution for implementing a common readout system for the iTOP and KLM upgrades after LS2.

Operating electronics within the Belle II detector volume presents numerous engineering challenges. The electronics must function not only in a harsh radiation environment \cite{belle_rad_env} but also within a strong longitudinal magnetic field of approximately 1.5 T \cite{belle_B_field}.  
To mitigate radiation effects such as Single Event Upset (SEU) and Single Event Latch (SEL), techniques for the RFSoC are provided by AMD UG584 user guide\cite{lvaux}. Data on operating RFSoC in such extremely high magnetic fields are unavailable, introducing a project risk when considering RFSoC as a viable solution for the Belle II upgrade. Before RFSoC can be deemed a viable solution as a high-speed digitizer for the Belle II upgrade post-LS2, validation is necessary to ensure that the RFSoC can operate without significant performance degradation. The objective of this paper is to determine whether the RFSoC can operate stably in a strong magnetic field and measure any performance degradation.

The establishment of RFSoC technology as a reliable high speed digitization platform under intense magnetic field conditions has implications that reach beyond the immediate needs of the Belle II experiment. Successfully demonstrating the RFSoC's performance and resilience in such an environment could guide the design of electronics for other high energy physics experiments, particularly those requiring compact, high frequency digitizers capable of enduring challenging operational conditions. Additionally, this technology could impact various fields facing similar constraints, such as medical imaging, which relies on precise electronic stability within magnetic fields. By showcasing a scalable, robust, and high performance solution, this study advances the broader field of resilient electronics, paving the way for adaptable, high fidelity digitization systems across a range of demanding scientific applications.


\section{Hardware}
\label{sec:hardware}

\begin{figure}[tb]
\centering
\includegraphics[width=1.0\textwidth]{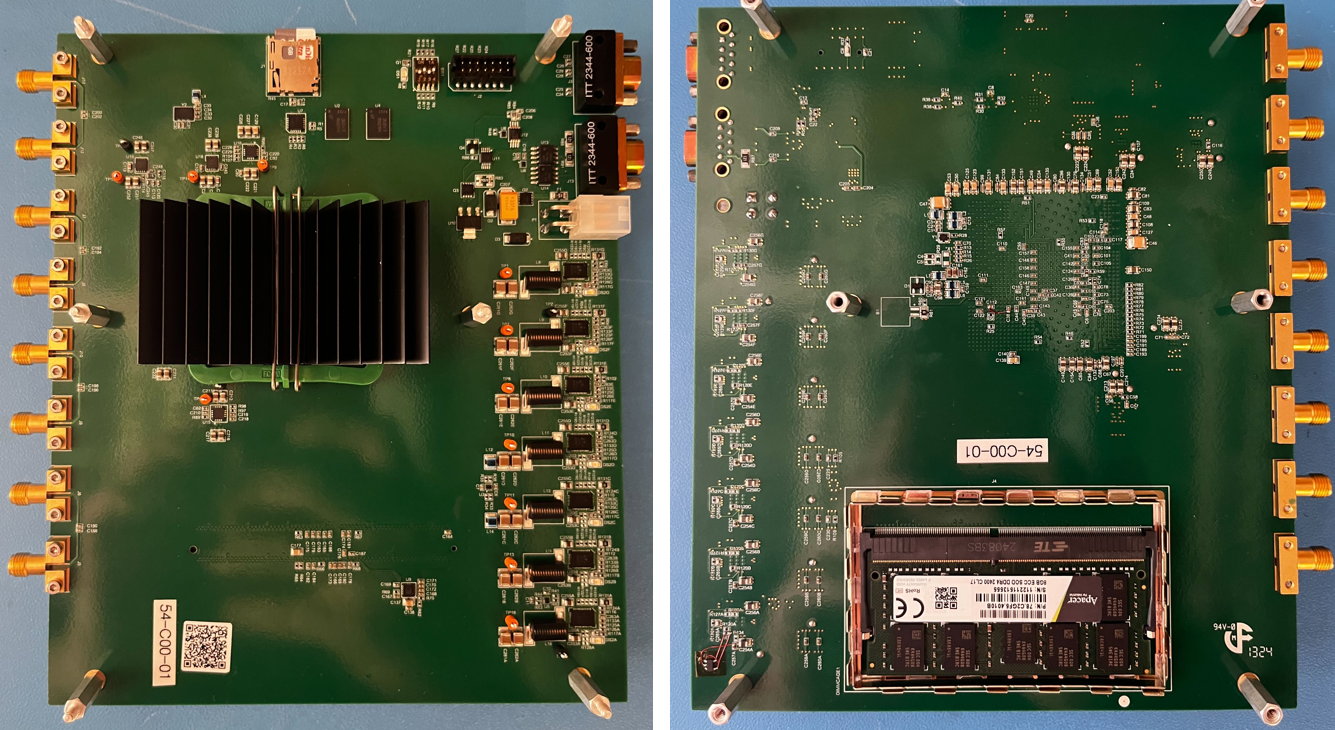}
\caption{\label{fig:hw_board_pic} Photograph of the custom PCB designed for strong magnetic field testing. The left image shows the top side of the board with the RFSoC located beneath the heatsink, while the right image displays the bottom side with the RFSoC's external memory EMI shield removed for clarity.}
\end{figure}

Ideally, this measurement would be performed by placing a commercially available design, such as a development board, into a strong magnetic field. However, finding a commercial development board free of magnetically sensitive components is highly unlikely due to the widespread use of iron-based inductors in DC/DC converters. To address this issue, a custom printed circuit board (PCB) was developed. This PCB features components around the RFSoC that are more resistant to external electromagnetic fields compared to standard commercial boards. Figure \ref{fig:hw_board_pic} shows a photograph of this custom PCB. The board's dimensions are 152.4 mm by 177.8 mm.

\begin{figure}[tb]
\centering
\includegraphics[width=1.0\textwidth]{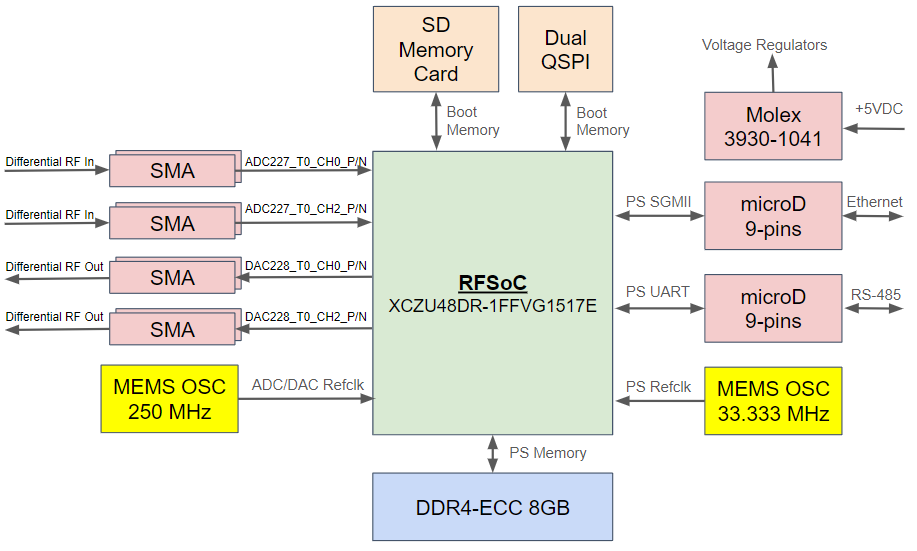}
\caption{\label{fig:hw_pcb_block_diagram} Block diagram of the custom PCB design}
\end{figure}

A block diagram of the custom PCB design is shown in Figure \ref{fig:hw_pcb_block_diagram}. The board is powered by a single 5 V power supply, providing power to all the DC/DC converters and linear voltage regulators. An Analog Devices LTC3815 IC, supporting an external inductor, is used for the DC/DC converters. For this DC/DC converter design, an air-core inductor (Abracon AIAC-4125C-R206J-T) was chosen for its superior immunity to external magnetic fields. Unlike iron-based inductors, air-core inductors lack a ferromagnetic core, which significantly reduces their susceptibility to magnetic saturation caused by strong external magnetic fields.

MicroElectroMechanical Systems (MEMS)-based reference oscillators for ADCs, DACs, and the RFSoC's Processor Side (PS) are used. MEMS oscillators were chosen for their superior immunity to strong external magnetic fields compared to standard quartz oscillators. This is because MEMS oscillators operate on electrostatic rather than electromagnetic principles, making them less susceptible to magnetic interference.

For the RFSoC's memory, an 8-GB DDR4 SODIMM with Error Correction Code (ECC) support is used. Two boot memory options are available: a dual Quad SPI (QSPI) and an SD memory card.

The RFSoC's Serial Gigabit Media Independent Interface (SGMII) connects to the Ethernet interface via a microD 9-pin connector, which enables the register configuration and streaming of ADC and DAC waveforms. Another microD 9-pin connector is used for the serial console interface to communicate with the RFSoC's Linux operating system.

Two of the eight differential ADC channels are connected to SMA connectors, while the other six channels are terminated with 50 Ohms (differential 100 Ohms) via the AC coupling capacitors. Similarly, two of the eight differential DAC channels are connected to SMA connectors, with the other six channels terminated with 50 Ohms (differential 100 Ohms) via the AC coupling capacitors. SMAs are used to provide flexibility for testing through cabling.

\begin{figure}[tb]
\centering
\includegraphics[width=0.80\textwidth]{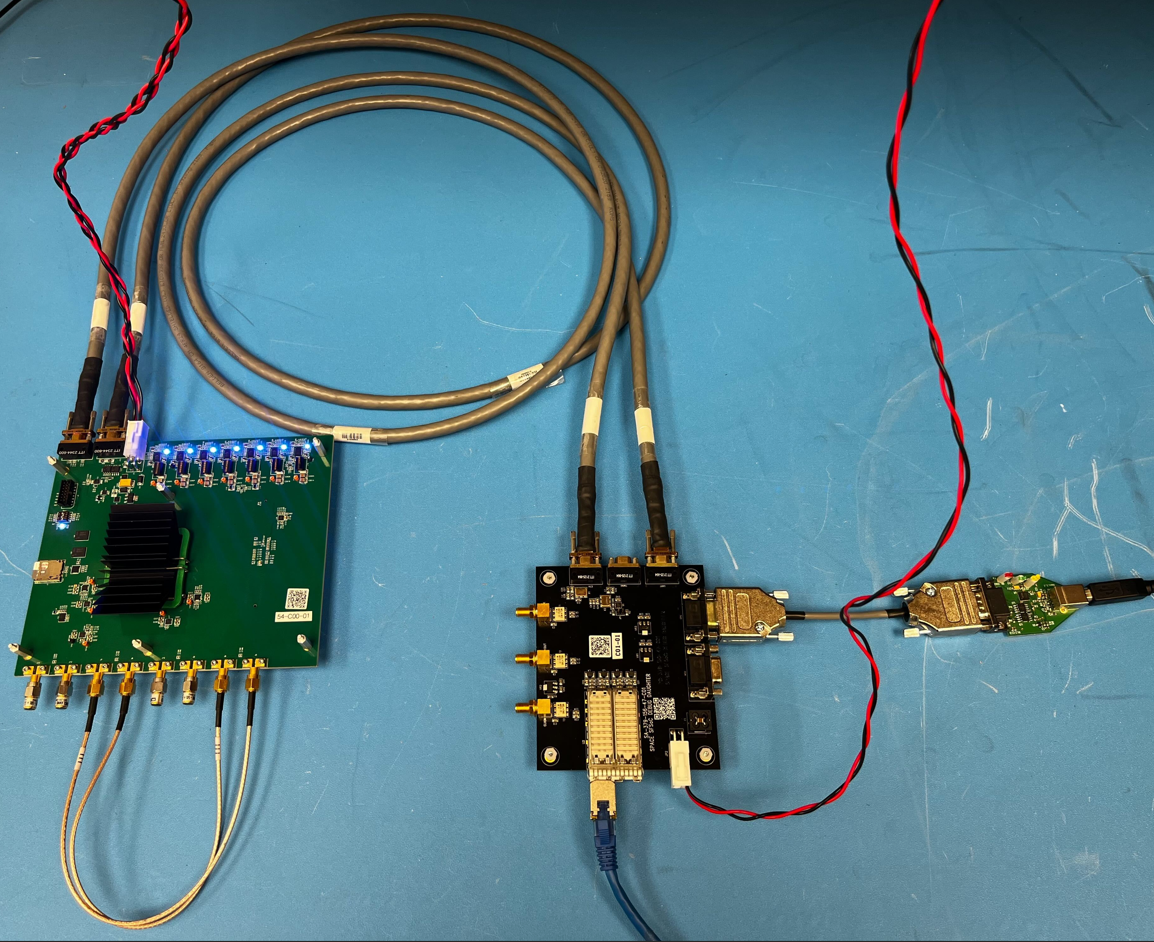}
\caption{\label{fig:hw_office_testing} Photograph of RFSoC board  on the left connected to a support board on the right}
\end{figure}

A photograph of the firmware and software development setup before the strong magnetic field testing is shown in Figure \ref{fig:hw_office_testing}. This photograph shows hardware setup used during the firmware and software development phase, prior to the actual strong external magnetic field testing. On the left, there is the custom RFSoC board. On the right, there is a custom support board, which converts the serial console interface (RS-485 protocol) to USB, and SGMII to a standard RJ45 Ethernet interface.

\section{Firmware}
\label{sec:firmware}
\begin{figure}[tb]
\centering
\includegraphics[width=1.0\textwidth]{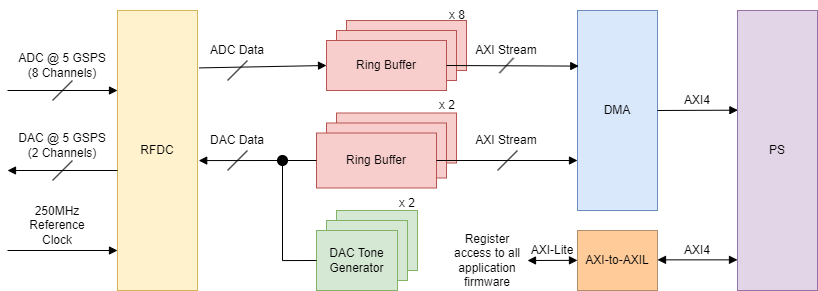}
\caption{\label{fig:fw_block_diagram} Block diagram of the firmware}
\end{figure}

A block diagram of the RFSoC firmware is shown in Figure \ref{fig:fw_block_diagram}. At the core of the digitization process is the Radio Frequency Data Converter (RFDC) block, a hardened Intellectual Property (IP) within the RFSoC responsible for ADC (Analog-to-Digital Converter) and DAC (Digital-to-Analog Converter) data conversion. The ADC is configured to operate at 5 GSPS across eight channels, receiving analog signals and converting them to digital data. Similarly, the DAC operates at 5 GSPS, converting digital data to analog signals across two channels. The on-board MEMS oscillator provides a 250 MHz reference clock. The RFDC's internal Phase-Locked Loops (PLLs) convert this 250 MHz reference clock into 5 GHz sampling clocks for the ADC and DAC channels.

Digital data from the ADC is stored in ring buffers, each with a recording time depth of 13 microseconds. There are eight buffers corresponding to the eight ADC channels. This high-speed data is temporarily stored and periodically read out at a slower rate (a few Hz per buffer). The data from the readout ring buffer is then transmitted as an AXI Stream to the Direct Memory Access (DMA) block via the AXI Stream protocol \cite{axis}. Subsequently, the DMA transfers the data over an AXI4 interface to the PS using the AXI4 memory protocol \cite{axil}.

The firmware includes two DAC Tone Generator modules, each capable of internally generating single-tone data for testing purposes. The data generated by these DAC Tone Generator modules is sent to both the RFDC and an additional set of ring buffers, with each buffer corresponding to one of the two DAC channels. 


Control and register access to all application firmware components is managed via an AXI-Lite interface \cite{axil}. A protocol conversion from the PS's native AXI4 memory interface to an AXI-Lite interface is used.

\section{Software}
\label{sec:software}

\begin{figure}[tb]
\centering
\includegraphics[width=1.0\textwidth]{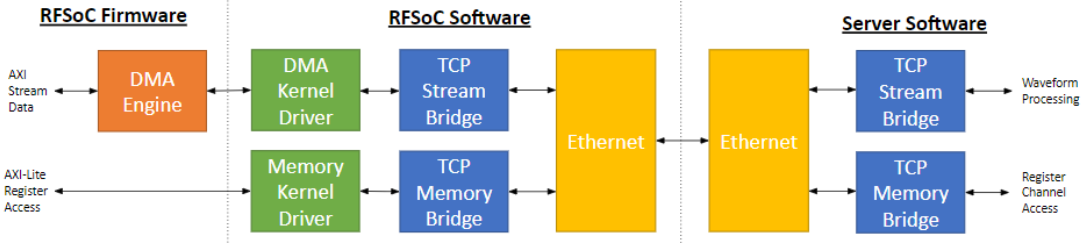}
\caption{\label{fig:sw_block_diagram} Block diagram of the software register access and waveform streaming.}
\end{figure}

In the architectural diagram shown in Figure \ref{fig:sw_block_diagram}, the integration of RFSoC firmware, RFSoC software, and server software is outlined. Within the RFSoC firmware domain, the waveforms are relayed to the DMA engine and the associated DMA kernel driver via the AXI Stream interface. Concurrently, AXI-Lite register access is performed through a memory kernel driver, which imposes restrictions on user access to registers outside of the designated application firmware register address space. A TCP stream bridge is used to transform the waveform data into TCP packets that are suitable for Ethernet transmission. Similarly, the TCP memory bridge serves to translate register access communications between the memory kernel driver and the Ethernet-based TCP interface. On the server side, the equivalent TCP stream and memory bridges work to encode and decode the TCP packets back into interfaces that are compatible with the native run control software. The RFSoC uses Petalinux (version 6.1.30-xilinx-v2023.2) to manage its Linux kernel operations.

\begin{figure}[tb]
\centering
\includegraphics[width=1.0\textwidth]{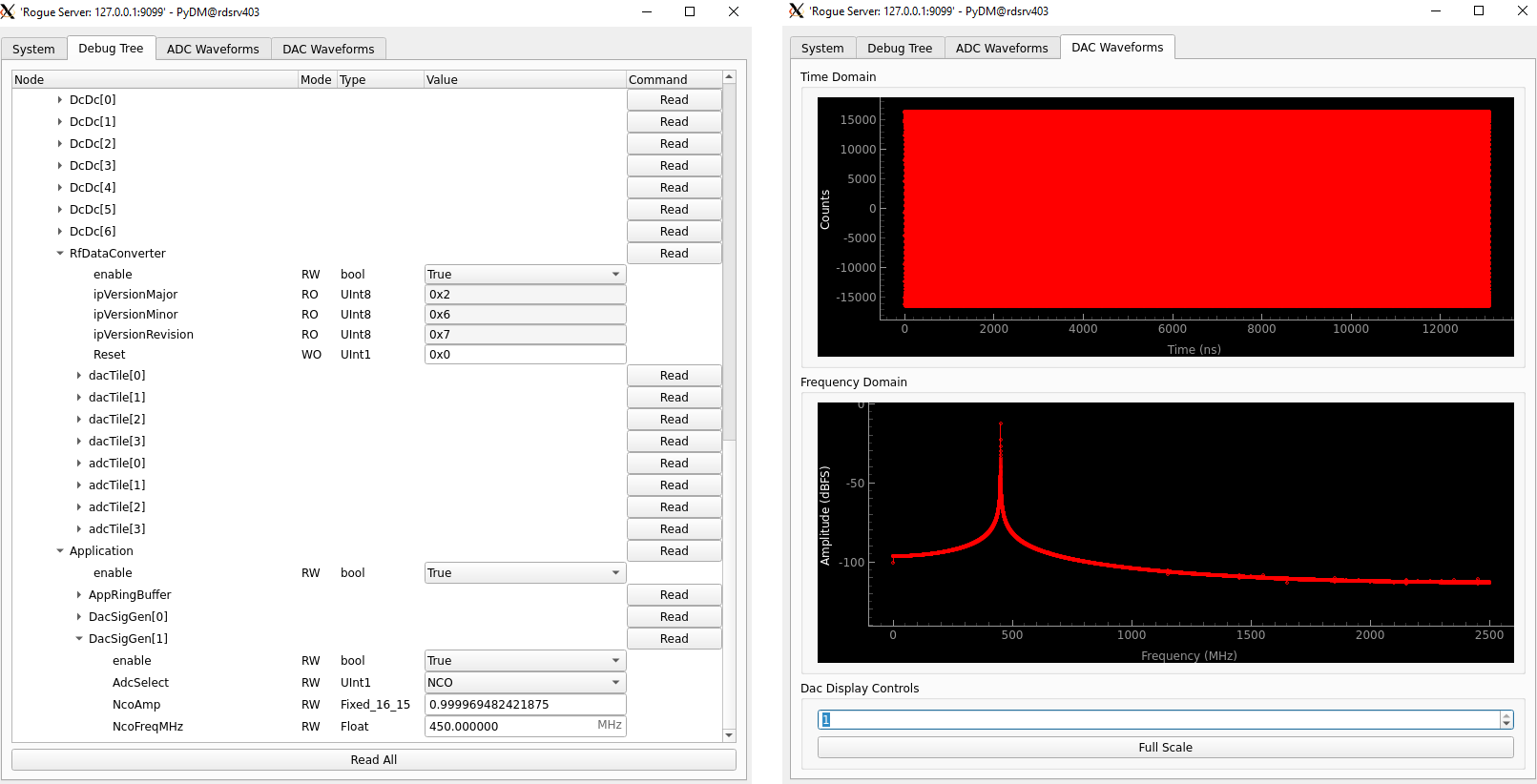}
\caption{\label{fig:sw_gui} Screenshots of the software GUI running on the rack server. (Left) Debug Tree for peak/poke of the RFSoC's registers, (Right) Live Display of the DAC waveforms.}
\end{figure}

The control software used in this testing, referred to as Rogue, is tailored for the swift prototyping and deployment of experimental setups. It supports execution in a hybrid Python/C++ mode or exclusively in C++. The software is compatible with x86-64, ARM32, and ARM64 platforms \cite{slaclab_rogue}. Rogue control software runs on both the RFSoC and the rack server, sharing a library for the TCP stream and memory bridges. Within Rogue, the application register mapping on the RFSoC is established through Python classes. Additionally, the system incorporates a Python class dedicated to transforming the raw byte array into a 16-bit numpy array, facilitating the analysis and processing of waveform data. A graphical user interface (GUI) has also been developed using the Rogue platform to enable live-display monitoring of ADC/DAC waveforms and registers, as shown in the example screenshot in Figure \ref{fig:sw_gui}. This GUI leverages the Python Display Manager (PyDM), which is based on PyQt, to create user interfaces for control systems \cite{slaclab_pydm}.

\section{Testing}
\label{sec:testing}

\subsection{Magnetic Field Setup}

\begin{figure}[tb]
\centering
\includegraphics[width=0.80\textwidth]{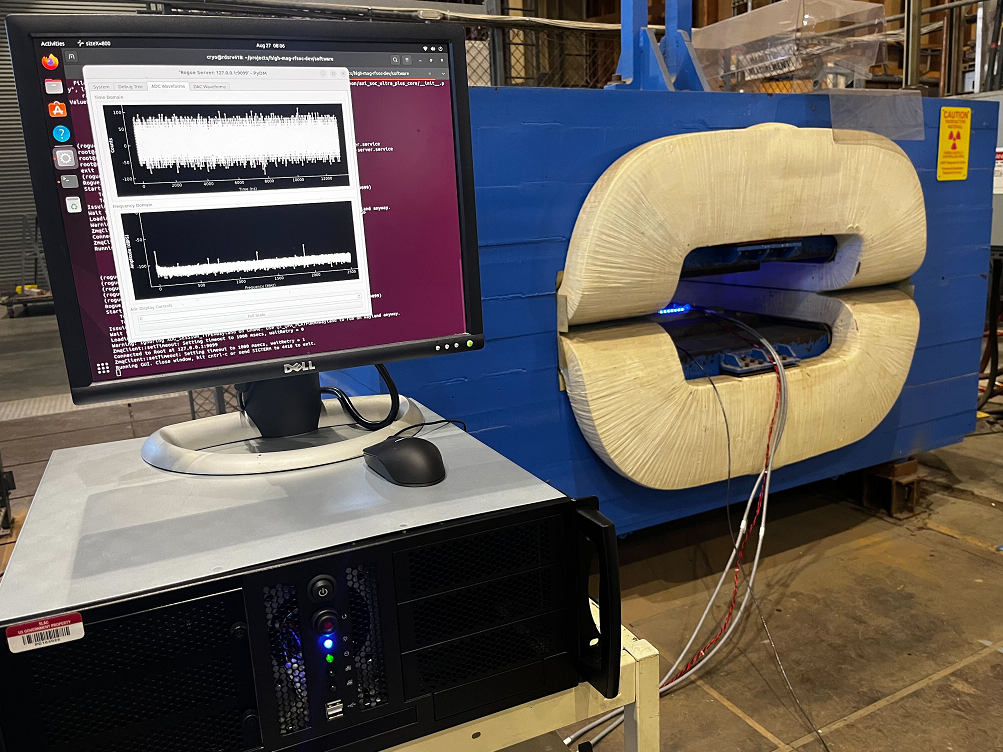}
\caption{\label{fig:testing_hw_setup} Photograph of RFSoC board running in the magnet on the right with the software GUI displayed on the left}
\end{figure}

A photograph of the test setup is shown in Figure \ref{fig:testing_hw_setup}. The magnet used, referred to as "Big Blue," is located at the SLAC National Accelerator Laboratory in the Heavy Fabrication Building (Building 026). The magnet measures 2.5 meters wide, 0.9 meters deep, and 1.3 meters high, with an approximate weight of 23 metric tons. It can generate a uniform magnetic field of up to 1.25 T in the center of the gap, where the RFSoC board was positioned. Although the testing was ideally meant to be conducted at a field strength of 1.5 T, matching the conditions within the Belle II detector, the SLAC magnet's maximum field strength was limited to 1.25 T due to power supply and cooling limitations.
The lower magnetic field strength of 1.25 T, compared to the target 1.5 T, may not fully replicate the potential performance impacts or failures that could occur in the RFSoC at the higher field strength within the Belle II detector. Therefore, additional testing at 1.5 T will be necessary to conclusively determine the RFSoC’s suitability for the Belle II upgrades.

The support board and server were connected via cables, positioned outside the magnet, in an area with minimal stray magnetic fields. During each magnetic field increment, waveforms were collected, and all status registers in the RFSoC were logged. Accurate magnetic field readings for each step were provided by a Group 3 DTM-141 Digital Teslameter module, with its probe placed in the magnet's gap.

\subsection{Loopback Testing in Magnetic Field}
\label{sec:testing_loopback}

\begin{figure}[tb]
\centering
\includegraphics[width=1.0\textwidth]{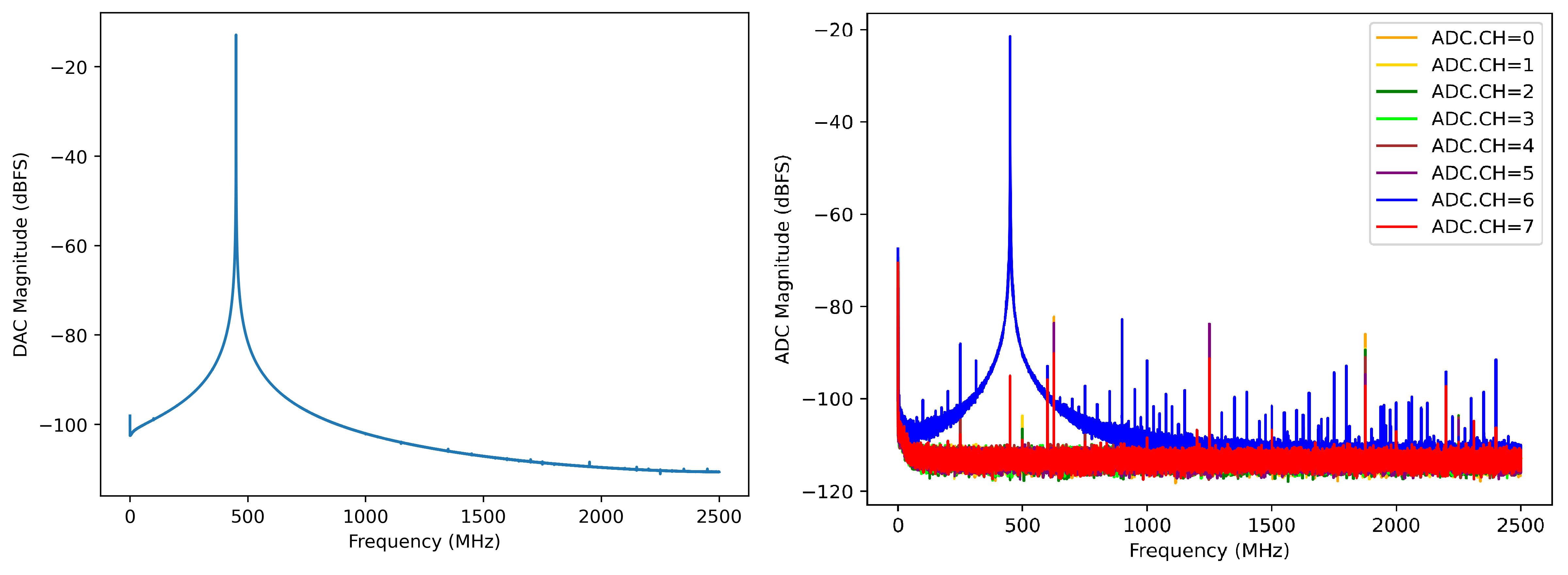}
\caption{\label{fig:testing_fft} 
(Left) FFT for DAC Channel 1, (Right) FFT of all ADC channels at while operating in 1.25 T field
}
\end{figure}

SMA loopback cables were connected from the differential pair of DAC channel 1 to the differential pair of ADC channel 6. DAC channel 0 and ADC channel 7 were terminated using 50 Ohm SMA terminators, while all other ADC and DAC channels were already terminated onboard. A 400 MHz single tone was generated on DAC channel 1 via firmware (refer to Figure \ref{fig:fw_nco}) and measured by ADC channel 6 to characterize the RFSoC digitization performance during magnet testing. The Fast Fourier Transform (FFT) plots for DAC channel 1 and all ADC channels at 1.25 T are shown in Figure \ref{fig:testing_fft}.

Spurious-Free Dynamic Range (SFDR) is an effective metric for characterizing ADC performance, as it measures the ADC's ability to distinguish the desired signal from the largest spurious signal above the noise floor. This metric reflects the ADC's precision and dynamic range in environments where spurious signals and noise can interfere with accurate signal representation. The SFDR is determined by calculating the difference between the power of the fundamental signal and that of the largest spurious signal. The largest spurious signal, observed at 800 MHz, was identified as the second harmonic of the fundamental tone on ADC channel 6. The next largest noise spurs, observed across all ADC channels and shown in Figure \ref{fig:testing_fft}, were located at 625 MHz, 1.25 GHz, and 1.875 GHz. These spurs are attributed to the 8-fold interleaved structure of the integrated ADCs in the RFSoC~\cite{PG269,rfsoc_chao}.

\begin{figure}[tb]
\centering
\includegraphics[width=0.75\textwidth]{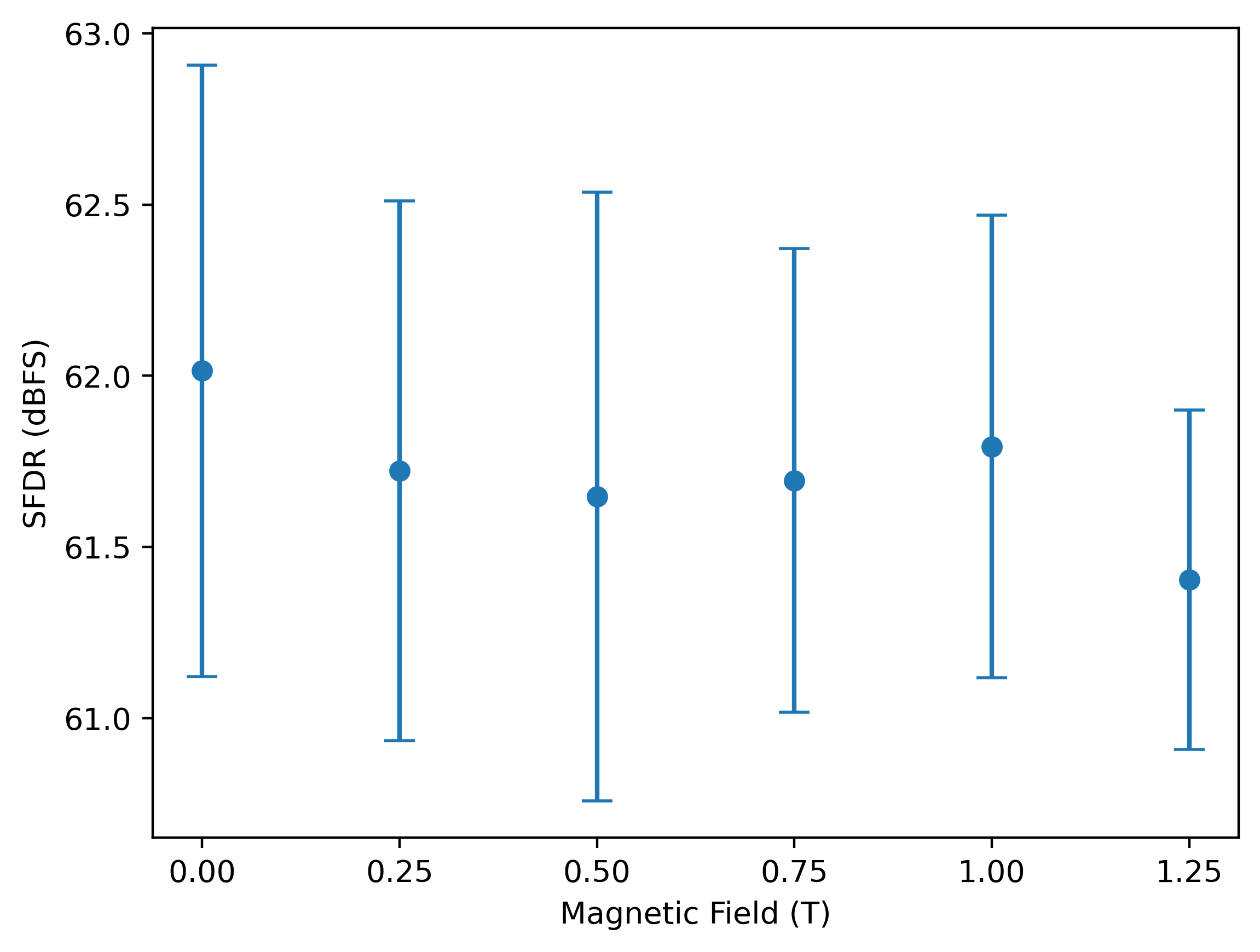}
\caption{\label{fig:SFDR_plot} 
Plot of SFDR versus magnetic field 
}
\end{figure}

ADC waveforms were collected at approximately 0.25 T magnetic field increments, with 100 waveforms recorded per channel per step. ADC channel 6, which received the single-tone signal from the DAC, was used to calculate the SFDR for each magnetic field step. A plot of SFDR versus magnetic field is presented in Figure \ref{fig:SFDR_plot}. The SFDR remained consistent across the entire magnetic field range, demonstrating that the ADC and DAC exhibit significant immunity to external strong magnetic fields.

\subsection{Non-orthogonal Magnetic Field Operation}

\begin{figure}[tb]
\centering
\includegraphics[width=1.0\textwidth]{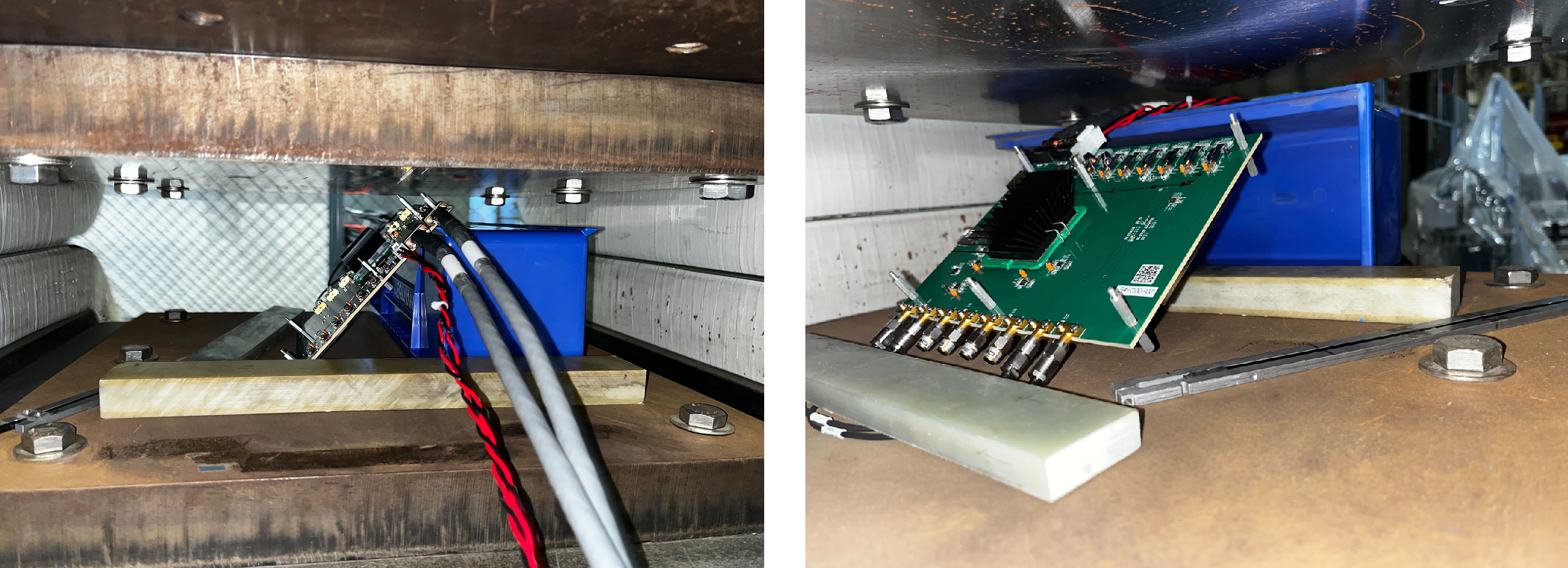}
\caption{\label{fig:test_pic_rotate} (Left) Photograph of RFSoC board rotated on X-axis, (Right) Photograph of the RFSoC board rotated on the Y-axis
}
\end{figure}

For the testing described in Section \ref{sec:testing_loopback}, the magnetic field was oriented orthogonally to the RFSoC and PCB. However, if the RFSoC is to be used in the Belle II detector, it will need to operate in non-orthogonal field configurations. Therefore, the RFSoC board was tested in two additional configurations at 1.25 T, as shown in Figure \ref{fig:test_pic_rotate}. In the first non-orthogonal configuration, the board was rotated about its X-axis by 51$^\circ$. In the second configuration, the board was rotated about its Y-axis by 42$^\circ$. Ideally, these rotations would have been 90$^\circ$, but the physical dimensions of the board and its cabling prevented this. 
Due to a miscommunication regarding the actual size of the magnet gap, it was mistakenly assumed that the board could be larger than the available space, leading to a PCB design that exceeded the intended dimensions for accommodating 90$^\circ$ orientations. As a result, the rotations achieved represent the maximum feasible configurations within the unexpectedly limited space.

\begin{figure}[tb]
\centering
\includegraphics[width=0.75\textwidth]{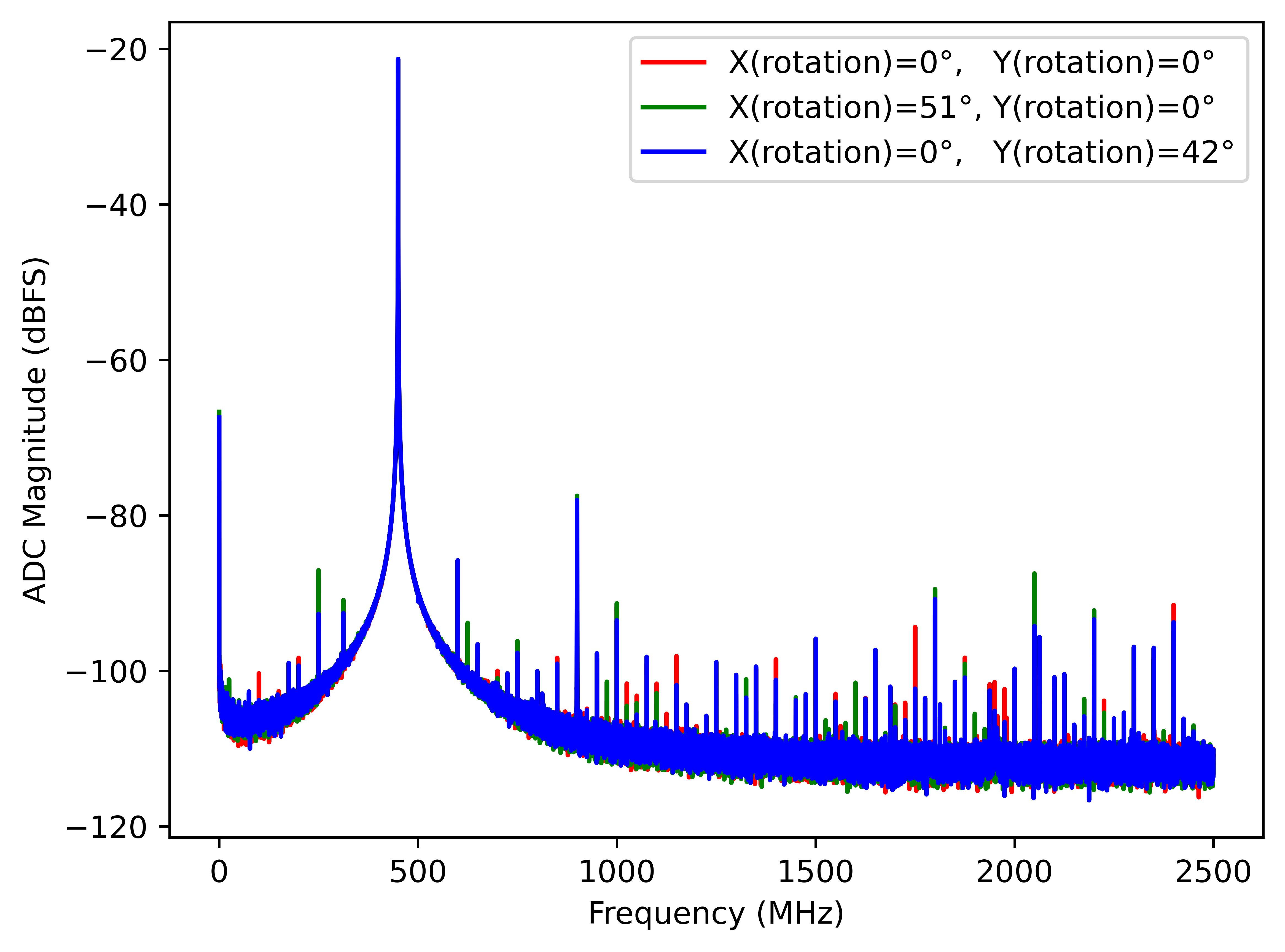}
\caption{\label{fig:Rotate_FFT_ADC} FFT of ADC channel 6 for the 3 different rotational configurations in a 1.25 T field
}
\end{figure}

For each of these non-orthogonal field configurations, 100 waveforms per channel were collected at 1.25 T, and compared with the orthogonal configuration described in Section \ref{sec:testing_loopback}. An FFT plot for ADC channel 6, overlaying the three configurations, is presented in Figure \ref{fig:Rotate_FFT_ADC}. An SFDR of approximately 62 dBFS was measured in both non-orthogonal field configurations, comparable to the SFDR observed in the orthogonal configuration. This indicates that the RFSoC has significant immunity to strong external magnetic fields, even at different incident angles of the magnetic field.

\subsection{RFSoC Power Draw Monitoring}

For the DC/DC converters, Analog Devices' LTC3815 ICs were used. This IC has a Power Management Bus (PMBus) interface that was connected to the RFSoC's I/O. This PMBus provides useful diagnostics on the DC/DC converters, such as voltage, current, and power for both the input and output of the voltage regulator, with approximately 1\% accuracy. After each magnet step and collecting ADC waveforms, a dump of all the firmware status registers was logged, including the PMBus registers, for offline analysis.

\begin{figure}[tb]
\centering
\includegraphics[width=1.0\textwidth]{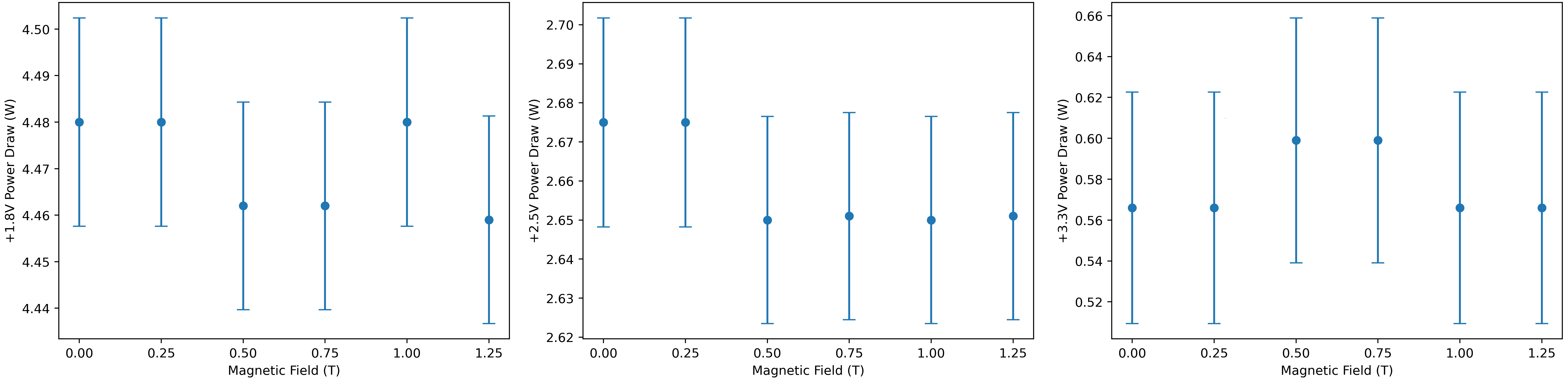}
\caption{\label{fig:testing_pwr_io}
(Left) Plot of +1.8V DC/DC converter's power output versus magnetic field, 
(Center) Plot of +2.5V DC/DC converter's power output versus magnetic field, 
(Right) Plot of +3.3V DC/DC converter's power output versus magnetic field 
}
\end{figure}

The DC/DC 1.8V voltage supply is regulated to 0.925V by linear voltage regulators for the RFSoC's analog core voltages: ADC\_AVCC and DAC\_AVCC. The DC/DC 2.5V voltage supply is regulated to 1.8V by linear voltage regulators for the RFSoC's analog auxiliary voltages: ADC\_AVCCAUX and DAC\_AVCCAUX. The DC/DC 3.3V voltage supply is regulated to 3.0V by a linear voltage regulator for the for the RFSoC's DAC output driver supply (DAC\_AVTT). As shown in Figure \ref{fig:testing_pwr_io}, the power drawn from these three rails, which supply the RFSoC's ADC and DAC circuits, remains fairly consistent across the entire tested magnetic field range. This consistency shows no evidence that the RFSoC's ADC and DAC circuits susceptible to external magnetic fields.

The power draw from the PL core voltage supply (Programmable Logic (VCCINT)) and the PS core voltage supply (Processor Side (VCCINT)), both regulated at 0.85V by separate DC/DC converters, are shown in Figure \ref{fig:testing_pwr_io}. While the PS VCCINT power draw remains fairly consistent across the entire tested range of magnetic fields, a linear dependency of the PL VCCINT power draw on strong external magnetic fields is evident. However, the increase in power draw is only about 5\% across this range of magnetic fields, which is minimal enough not to have a significant impact on the operation of the RFSoC in strong magnetic fields.

The observed linear dependency of the PL VCCINT power draw on the magnetic field can be attributed to several factors. One possible cause is increased noise or fluctuations in the power supply due to the magnetic field's effect on the on-board switching supplies. While air-core inductors experience smaller changes in inductance compared to those with ferromagnetic cores, these changes can still be noticeable in strong magnetic fields. This may lead to increased dynamic power consumption as the internal circuitry compensates for variations in the supply voltage. Another possibility is that magnetic fields could affect the behavior of the FPGA's internal PLLs and clock distribution circuits. Although these potential effects are relatively minor, they could collectively contribute to the approximately 5\% increase in PL VCCINT power draw. Nevertheless, the magnitude of this power change is small enough not to significantly impact the operation of the RFSoC in strong magnetic fields.

\begin{figure}[tb]
\centering
\includegraphics[width=1.0\textwidth]{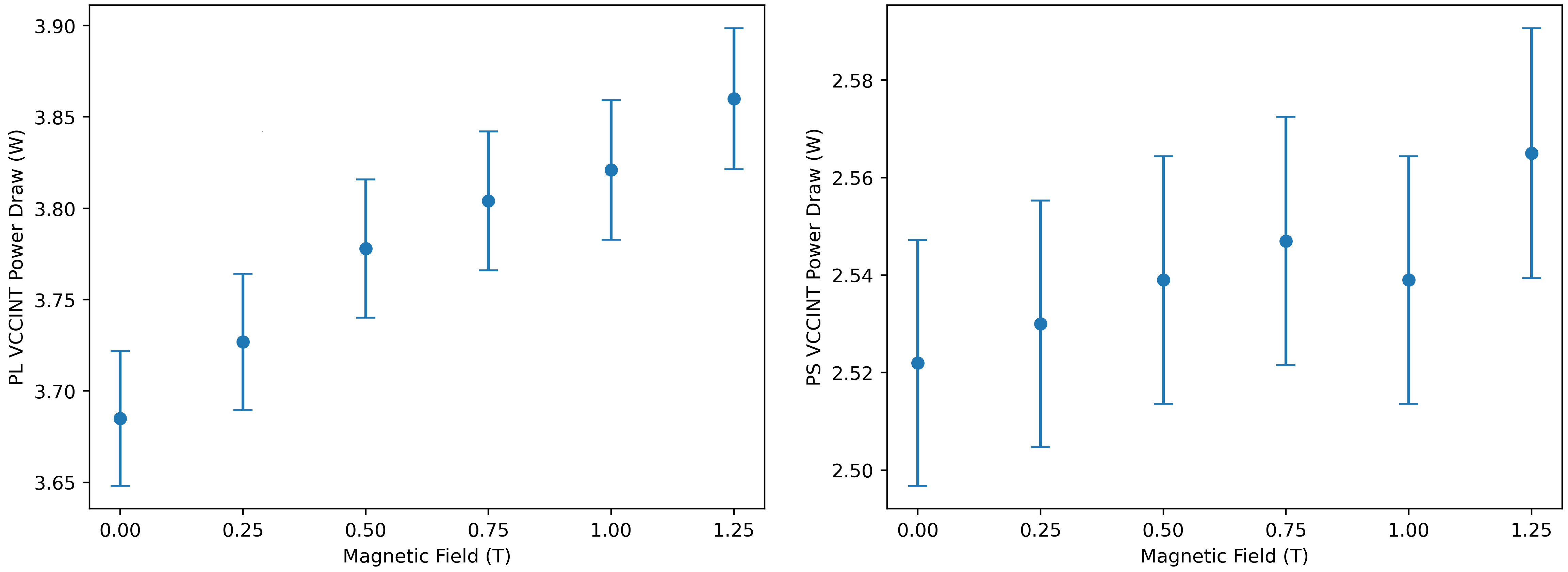}
\caption{\label{fig:testing_pwr_vccint}
(Left) Plot of PL VCCINT DC/DC converter's power output versus magnetic field, 
(Right) Plot of PS VCCINT DC/DC converter's power output versus magnetic field
}
\end{figure}

\subsection{Boot Memory Testing}

With a magnetic field set at 1.25 T and the RFSoC initially in the power-down state, it was powered up using the SD memory card to boot. No software errors were observed in the PetaLinux boot logs of the RFSoC. No issues were observed when running the software using the Rogue software GUI, and the live waveforms displayed appeared similar to those from previous testing, with an SFDR of approximately 62 dBFS.

Subsequently, both the magnet and the RFSoC were powered down to change the DIP switch on the RFSoC board for QSPI boot. Once the magnetic field was ramped up to 1.25 T again, the RFSoC board was powered on for QSPI boot. Similar to the behavior observed during the boot from the SD memory, no software errors were noted in the boot log, and the waveform behavior was similar to that observed previously.

\section{Summary}
\label{sec:Summary}

The RFSoC was identified as a potential solution for the Belle II detector's upgrade, which requires its operation in a high magnetic field environment. To assess its viability, a custom RFSoC board was designed with peripheral components selected for their resistance to magnetic susceptibility, including air-core inductors and MEMS-based oscillators. Using this custom board, the performance of an RFSoC operating within a 1.25 T magnetic field was evaluated.



The results indicate that the RFSoC can operate reliably within a 1.25 T magnetic field without significant performance degradation, which is promising for its potential use in Belle II. Ideally, this evaluation would have been conducted in a 1.5 T field, since that is the magnetic field strength used in the Belle II detector volume. However, the positive results obtained at 1.25 T suggest that the RFSoC is likely to operate at 1.5 T as well. Further testing at 1.5 T will be necessary to conclusively confirm the RFSoC’s suitability for the Belle II upgrades.

The findings from this study underscore the potential of RFSoC technology as a resilient, high speed digitization solution capable of functioning reliably in intense magnetic fields, which bodes well for its integration into the Belle II detector upgrade. Beyond its application in Belle II, the demonstrated durability and precision of the RFSoC in strong magnetic field conditions can influence the design of other high-energy physics experiments that require compact, high-frequency digitizers in similarly harsh environments. Additionally, the success of this technology in our testing suggests potential benefits for other fields, such as medical imaging, where electronic stability in magnetic fields is essential.

\section{Acknowledgements}

Work supported by the U.S. Department of Energy, under contract number DE-AC02-76SF00515. Also special thanks to Scott Anderson at SLAC National Accelerator Laboratory for his invaluable assistance in operating the magnet during the testing process.


\end{document}